\documentclass[aps,prl,twocolumn,superscriptaddress]{revtex4-1}
\usepackage{graphicx}
\usepackage{amsmath}
\usepackage{color}
\usepackage[colorlinks=true, linkcolor=black, urlcolor=blue, citecolor=blue]{hyperref}

\DeclareMathOperator{\arcsinh}{arcsinh}

\begin{document}

\title{Phase diagram of thick ribbons in a bad solvent}

\author{Thanh-Son Nguyen}
\affiliation{Institute of Physics, Vietnam Academy of Science and Technology, 
10 Dao Tan, Ba Dinh, Hanoi, Vietnam}
\affiliation{Liquid Crystal Institute, Kent State University, Kent, Ohio 44242, USA}

\author{Jayanth R. Banavar}
\affiliation{Department of Physics, University of Maryland, College Park,
Maryland 20742, USA}

\author{Amos Maritan}
\affiliation{Dipartimento di Fisica, Universit\`a di Padova, CNISM and INFN,
Via Marzolo 8, I-35131 Padova, Italy}

\author{Trinh X. Hoang}
\email{hoang@iop.vast.vn}
\affiliation{Institute of Physics, Vietnam Academy of Science and Technology, 
10 Dao Tan, Ba Dinh, Hanoi, Vietnam}

\begin{abstract}
Ribbons are topological objects of biological and technological importance.
Here, we study the folding of thick ribbons with hydrophobic surfaces in a bad
solvent in regimes in which either the ribbon's thickness or the solvent molecule
size is not vanishingly small compared to the ribbon's width. Extensive Monte
Carlo simulations show that ribbons of various lengths and with a small
stiffness adopt several distinct configurations as the ground state that
include rolled (Archimedean spiral), curled, twisted and globule conformations.
Analytic and numerical calculations based on the consideration of putative
ground states lead to phase diagrams that qualitatively agree with the
simulation results. A symmetry breaking of the planar rolled configuration
in favor of the elongated twisted and the globular ribbons is observed on
increasing the solvent size.  Interestingly, the twisted ribbon is found as the
ground state in the absence of any energetic preference for twisting.  We show
that the twist of the DNA double helix structure can be stabilized when modeled
as a hydrophobic thick ribbon even in the limit of vanishing solvent size.
\end{abstract}

\maketitle

A ribbon is an intermediate object \cite{Mahadevan2010} between a polymer
\cite{deGennes} and a membrane \cite{Nelson} and is frequently encountered in
everyday life, in biology and in nanotechnology. As with a membrane, the
surface of a ribbon
provides a significant contact area to another surface, while its one
dimensional character is useful for a variety of purposes such as taping,
knotting, tying and decoration. In molecular biology, the DNA double helix
structure has been seen as a twisted ribbon \cite{Boudaoud} with its
sugar-phosphate backbones represented by the ribbon's edges. Notably, a certain
phase difference between the two twisting edges of the ribbon is associated
with the appearance of the DNA minor and major grooves \cite{BrandonTooze}. In
the realm of supramolecular structures, amyloid filaments \cite{Chiti2006}
predominantly comprised of protein $\beta$-sheet structures \cite{BrandonTooze}
have the symmetry of a ribbon. 
Other notable examples include helical and twisted ribbons self-assembled by
amphiphilic molecules \cite{MacKintosh,Ziserman} and nanoribbons made up of a
single layer of atoms, such as graphene \cite{GeimRMP}.

Owing to both its 1D and 2D character, a ribbon can display distinct behaviors
under various conditions \cite{Mahadevan2010,Nelson2016}. Gauss's {\it Theorema
Egregium} shows that a flat ribbon can be bent into a helical (or spiral)
configuration with cylindrical symmetry preserving its zero Gaussian curvature,
but cannot be isometrically transformed into a twisted configuration of
negative Gaussian curvature. Twisting a flat ribbon requires non-isometric
deformations and may result in an increase of in-plane elastic energy
\cite{Ghafouri}. There has been interest on the transition
between helical and twisted ribbons \cite{MacKintosh,Ziserman}. 
Selinger {\it et al.} have shown that such transition is smooth for ribbons
formed by chiral molecules \cite{Selinger2004}. Ghafouri and Bruinsma have
argued that this transition becomes discontinuous on varying the ribbon's width
\cite{Ghafouri}.  Hatwalne and Muthukumar have recently suggested that
topological surface defects may contribute to the twisted geometry
\cite{Muthukumar2010}.
In these studies, chiral configurations arise from the interplay between
complex elastic forces and intrinsic molecular or material properties.

\begin{figure}
\includegraphics[width=\columnwidth]{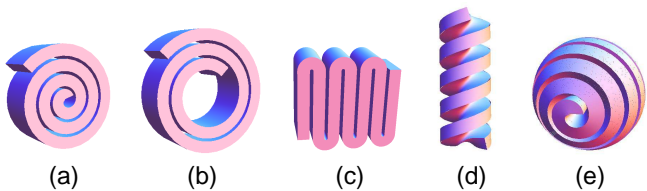}
\caption{
Various configurations of a thick ribbon including a rolled (Archimedean
spiral) conformation with no hole (a), a rolled conformation with a hole (b), a
curled (c), a twisted (d) and a globular (spherical spiral) (e) conformation. 
}
\label{fig:cartoon}
\end{figure}

In this Letter, we present a new and simple mechanism that leads to the
formation of a number of characteristic conformations of a ribbon such as the
rolled (Archimedean spiral \cite{Kreyszig}), the curled, the twisted and the
globular conformations (Fig. \ref{fig:cartoon}).  This mechanism does not rely
on a ribbon's specific elastic properties, but rather on the geometry of
the ribbon and the grained character of the surrounding solvent. In particular,
we consider a hydrophobic thick ribbon in solution and study how the ribbon
thickness and the solvent molecule's size may impact the ribbon's phase
behavior. We will show that both the ribbon thickness and the solvent size give
rise to the formation of the twisted conformation albeit the ribbon has no
intrinsic elastic chiral preference. As an application, we will show how the
model of hydrophobic thick ribbon can be tailored to stabilize the twist of DNA
molecule.

The geometric approach employed here for the thick ribbons follows from
earlier work used to model thick polymers or flexible tubes
\cite{MaritanNature,BanavarAnnRev07,Kamien,Dietrich}. It has been shown that a
hydrophobic tube curls into a tight helix or an almost planar sheet when
immersed in water \cite{Banavar2007}. Interestingly, these emergent motifs form
the building blocks of protein structures \cite{BrandonTooze}.  We will
consider two driving forces for the folding of a ribbon: we seek to minimize the
ribbon's exposed area to the solvent while minimizing its the bending energy.
Based on extensive computer simulations under various conditions, we find that
the ribbon commonly forms just a few optimal conformations similar to the ones
shown in Fig. \ref{fig:cartoon}. Armed with this insight, we have carried out
analytic and numerical calculations of the energies of these putative ground
states and elucidate a phase diagram. These calculations are vastly simpler
than doing the simulations -- ensuring that a ribbon of non-zero thickness does
not self-intersect requires a set of 4-body non-local interactions
\cite{BanavarJSP,Hoang2012EPL}, which are highly computer intensive.  As is
common in such optimization problems, one cannot definitively rule out the
existence of other novel ground states that we have not encountered in our
simulations. However, our simulations suggest that the phase diagram we have
obtained is in fact complete. 

Consider a ribbon of length $L$, width $W$ and thickness $h$. We will assume
that the two main surfaces associated with the width of the ribbon (the upper
and lower surfaces) are hydrophobic to the solvent, whereas the side surfaces
associated with its thickness are neutral. Solvent molecules are considered as
spheres of diameter $D$. The ribbon is considered to be flexible but its
midplane surface area, width and thickness are conserved.  
The total energy of the ribbon (we work exclusively in the low temperature limit
and do not consider entropic effects) is given by \cite{Dietrich,Hadwiger}
\begin{equation}
F = \sigma S + U \ ,
\label{eq:F}
\end{equation}
where $S$ is the ribbon's total exposed surface area, $\sigma > 0$ is the
solvent-induced energy per unit surface area, and $U$ is the bending
energy. $S$ is defined on both upper and lower surfaces, whereas for
simplicity, $U$ is considered only for the ribbon's midplane.  For the latter,
we employ the Helfrich's free energy density of curvature \cite{Helfrich} 
\begin{equation} \label{fc}
f_\mathrm{c} = \frac{\kappa}{2} H^2 + \bar\kappa\, K \ ,
\end{equation}
where 
$H=c_1+c_2$ and $K=c_1
c_2$ are the mean and Gaussian curvatures, respectively, with $c_1$, $c_2$ the
two principal curvatures of the surface \cite{Kreyszig}; $\kappa$ and
$\bar\kappa$ are bending rigidities. 
Note that a positive $\bar\kappa$ would make the twisted conformation
elastically favorable because of its negative Gaussian curvature.
In this study, we consider however a special case of $\bar\kappa = - \kappa <
0$, for which the Helfrich's free energy density reduces to a simpler form,
\begin{equation}\label{fcc}
f_\mathrm{c} = \frac{\kappa}{2} (c_1^2 + c_2^2) \ ,
\end{equation}
with $\kappa$ measured in units of $\sigma W^2$.  This free energy density
entails no energetic preference for both the bending and the twisting of the
ribbon.  The total bending energy $U$ is obtained by integrating $f_\mathrm{c}$
over the midplane. Note that the present model does not include in-plane
elastic energy.

We first carried out Monte Carlo simulations of hydrophobic thick ribbons to
find their ground states.  A model of ribbon appropriately adapted to the
general description given above is necessary for the simulations.  The ribbon's
midplane is represented by spherical beads on a square mesh. The beads along
the ribbon's lateral dimension are always in a straight line with a fixed
lattice spacing equal to $b$, so that the ribbon can be considered as a ladder
made up of straight rungs. The bead spacing along the ribbon's central curve
is also fixed and equal to $b$ whereas on the rest of the lattice it is allowed
to vary between $0.5b$ and $1.5b$ thereby permitting the ribbon to bend and
twist. Self-avoidance \cite{BanavarJSP,Hoang2012EPL} requires that the radius
of a sphere which is tangent to the ribbon at a given bead (in either the
normal or the anti-normal direction) and passing through any of the other
beads ought to be larger than $h/2$. Additionally, a hardcore radius of $b/4$
is introduced for every bead to avoid spurious overlap, given that $h >
b/2$. A bending energy is applied for the beads on the central curve and is
given by
\begin{equation}
u = \kappa_b (1 - \cos \theta) + \kappa_b (1 - \cos \phi) ,
\end{equation}
where $\kappa_b = \kappa (b/W)^2$ is the stiffness per bead, the first
term on the r.h.s.  is the bending energy with $\theta$ denoting the bond
(bending) angle at the
bead and the
second term is the twisting energy with $\phi$ representing the angle between
two consecutive rungs at the given site. It is also assumed that all the beads
on the same rung have the same bending energy. The ribbon exposed area is
calculated by counting the number of surface elements exposed to the solvent.
In order to check if an element is exposed to the solvent, one places a solvent
sphere of diameter $D$ centered at a distance of $(h + D)/2$ from a midplane
bead in the normal or anti-normal direction and then checks whether it is found
at a distance larger than $(h + D)/2$ from all other beads. The samplings are
carried out by parallel tempering \cite{Swendsen} simulations with pivot
and reptation \cite{deGennes} moves. 

\begin{figure}
\includegraphics[width=\columnwidth]{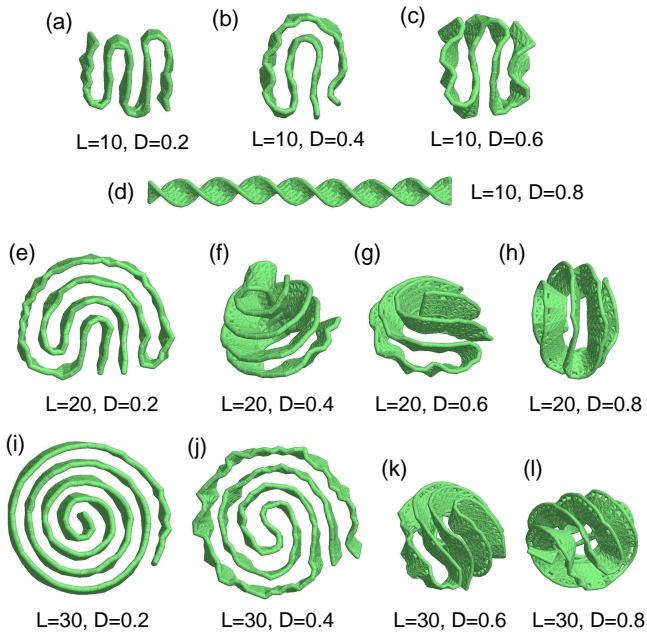}
\caption{
Lowest energy conformations (a-l) of thick ribbons obtained by simulations for
various lengths $L$ and solvent radii $D$ as indicated. The ribbon parameters
used in the simulations are: $W=1$ (width), $h=0.5$ (thickness), $b=0.2$ (bead
spacing), and $\kappa_b=0.1\,\sigma b^2$ (equivalent to $\kappa=0.1\,\sigma
W^2$).
}
\label{fig:btwist2}
\end{figure}

Fig. \ref{fig:btwist2} shows the ground states obtained by the simulations for
ribbons of several lengths and for different solvent sizes. It is shown that
for small solvent diameter ($D=0.2W$), the ground state has a planar symmetry
which can be either a curled ($L=10W$), a rolled ($L=30W$) or some intermediate
conformation ($L=20W$). As the solvent size increases, the planar symmetry
is broken -- the ground state becomes a twisted conformation for the smallest
$L$ or some sorts of globular conformation for larger $L$. 

\begin{figure}
\includegraphics[width=\columnwidth]{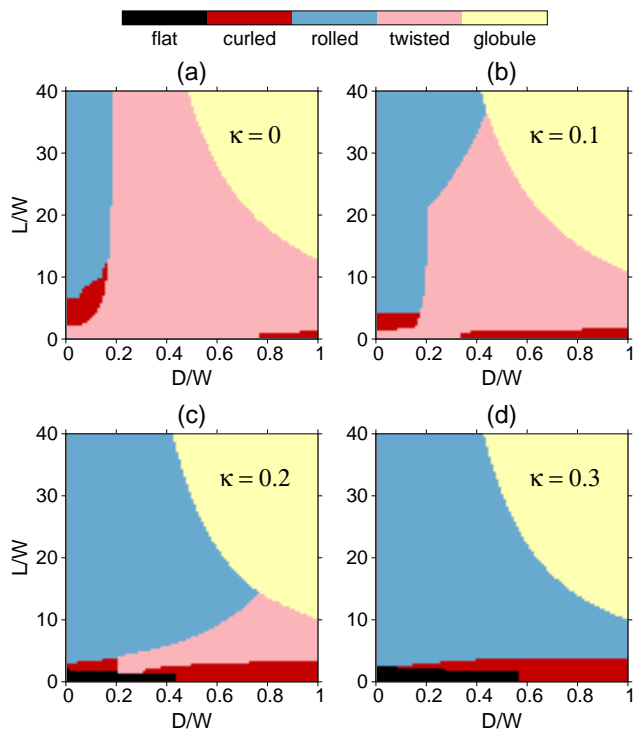}
\caption{
Ground state phase diagram as function of the solvent diameter $D$ and the
ribbon's length $L$ given in units of the ribbon's width $W$.  The ribbon's
thickness is $h=0.5W$.  The phase diagrams were obtained for several values of
bending stiffness, $\kappa=0$ (a), $0.1\,\sigma W^2$ (b), $0.2\,\sigma W^2$ (c)
and 0.3 $\sigma W^2$ (d). Different phases are indicated by colors
as given by the legends (top).} 
\label{fig:diagram}
\end{figure}

We have analytically and numerically determined the phase diagram of a
hydrophobic thick ribbon by considering the rolled, the curled, the twisted and
the globular (spherical spiral) conformations as shown in Fig.
\ref{fig:cartoon} as candidates for the ground state. All these conformations
are parameterizable and thus their exposed area and bending energy can be
precisely calculated by using differential geometry and numerical methods (see
Supplemental Material \cite{SM}). For example, the midplane of the twisted
ribbon is considered as an ideal helicoid \cite{Kreyszig} whose Cartesian
coordinates are given by 
$\vec R_\mathrm{mid} (u,z) = (u \cos kz, u \sin kz, z)$, where $u \in
[-W/2,W/2]$ is the parameter of the ribbon lateral dimension and $k$ is the
wave number of the twist along the $z$ axis. The physical surfaces of the
ribbon then are constructed from the midplane. For a twisted ribbon of finite
thickness $h$, self-avoidance imposes that $k \leq 2/h$, and thus prevents
infinite twisting of the ribbon.  On the other hand, the surface of a twisted
ribbon is fully exposed to a solvent molecule of diameter $D$ if 
\begin{equation}
h+D \leq \frac{2}{k}\ ,
\end{equation}
otherwise it can become partially or fully buried on increasing $k$. An
interesting property of the twisted thick ribbon, that is due to the non-zero
thickness, is that the ribbon total hydrophobic surface area decreases with
$k$. Such property is not present for the rolled and curled ribbons.

The asymptotic behaviors of the ribbon exposed area and bending energy 
in the putative conformations at large $L$ can be analyzed. It is found that
for compact folding
\begin{eqnarray}
S_\mathrm{rolled} \propto h^{1/2} L^{1/2} & \qquad & 
U_\mathrm{rolled} \propto h^{-1} \ln L \\
S_\mathrm{curled} \propto h^{1/2} L^{1/2} & & 
U_\mathrm{curled} \propto h^{-3/2} L^{1/2} \\
S_\mathrm{twisted} \propto L (0) &  & 
U_\mathrm{twisted} \propto L  \\
S_\mathrm{globule} \propto L (0) & & 
U_\mathrm{globule} \propto h^{-1} \ln L \ ,
\end{eqnarray}
where the subscript denotes the conformation type, and the width $W$ of the
ribbon, considered as the length unit, has been absorbed. 
$S_\mathrm{twisted}$ and $S_\mathrm{globule}$ are proportional to $L$
for small $D$, and equal to zero for sufficiently large $D$.
These scaling
behaviors indicate that, for large $L$, the ground
state must be either the rolled conformation or the globule because of
the $\ln(L)$ dependence of their bending energy.

Fig. \ref{fig:diagram} shows the ground state phase diagram as a function of
solvent diameter $D$ and length $L$ for a ribbon of thickness $h=0.5W$ with
several values of stiffness $\kappa$.  For a given $L$ and $D$, the 
energies of the putative ribbon conformations are minimized with respect to
their geometrical parameters to identify the ground state.  
The curled ribbon appears as the ground state only for small $L$,
typically smaller than $10W$.  For larger $L$, the phase diagram is dominated
either by the rolled at small $D$, or the twisted and the globular
conformations at larger $D$.  Note that the twisted phase is present only at
sufficiently low stiffness, i.e.  the cases of $\kappa < 0.3\, \sigma W^2$ in
Fig. \ref{fig:diagram}. As $\kappa$ increases, the twisted phase shrinks being
replaced by the rolled phase and eventually disappears at high stiffness (Fig.
\ref{fig:diagram}d).  On the other hand, for a given $\kappa$, the twisted
phase also disappears on decreasing the ribbon thickness (Fig.
\ref{fig:phaseh}, Supplemental Material).

\begin{figure}
\includegraphics[width=\columnwidth]{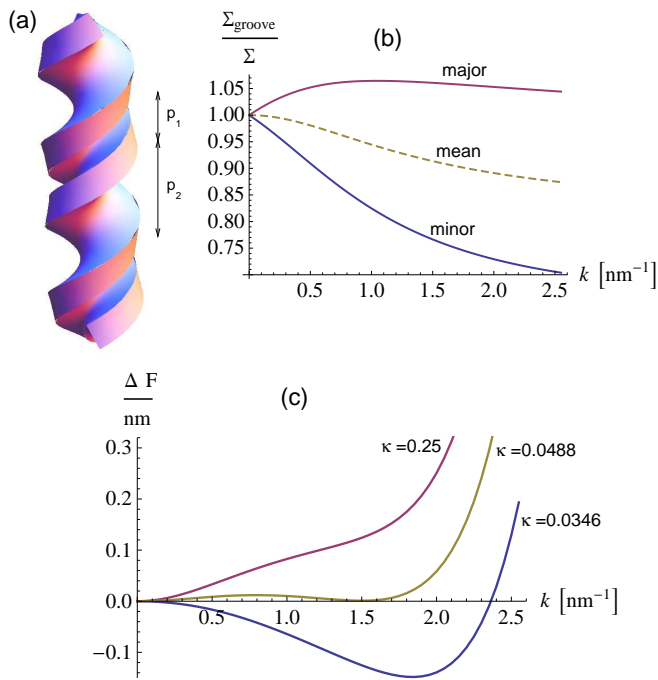}
\caption{(a) Thick ribbon representation of DNA molecule obtained with
realistic parameters for the B-DNA double helix structure (see text). The minor
and major grooves are indicated with their widths, $p_1$ and $p_2$,
respectively.  (b) Dependence of the groove surface area,
$\Sigma_\mathrm{groove}$, relative to that of the midplane, $\Sigma$, on the
wave number $k$ of DNA twist.  Dashed line corresponds to the mean area of the
two grooves.  (c) The energy difference $\Delta F$ per unit length between
a twisted and a flat thick ribbon as a function of $k$, obtained with $D=0$ for
three values of $\kappa$, equal to 0.25, 0.0488 and 0.0346 $\sigma W^2$, as
indicated.  For the last value of $\kappa$, the minimum of $\Delta F$
corresponds to the pitch of B-DNA, $p=3.4$ nm.
}
\label{fig:dna}
\end{figure}

The DNA molecule can be considered as a thick ribbon of width $W \approx 2$ nm
and thickness $h \approx 0.6$ nm with the known hydrophobicity
of its interstrand surfaces \cite{BrandonTooze}. The DNA twist in its
double helix structure, however, is not an ideal helicoid as it bends and winds
differently on its two main surfaces leading to the formation of major and
minor grooves. We parameterize the midplane of the DNA ribbon as
\begin{equation}
{\vec R}_\mathrm{mid} (u,z) = (u \cos \delta \cos kz, u \cos \delta \sin kz, z 
+ u \sin \delta) ,
\end{equation}
where $u \in [-W/2,W/2]$, $k=2\pi/p$ is the wave vector of the twist with $p$
the helical pitch, $\delta$ is the tilt angle of the ribbon lateral direction
from the plane perpendicular to the twist axis ($\delta=0$ for an ideal
helicoid). For B-DNA, $p=3.4$ nm is the sum of the widths, $p_1=1.2$ nm and
$p_2=2.2$ nm, of the minor and major grooves, respectively (see Fig.
\ref{fig:dna}a). The tilt angle can be calculated as $\sin \delta = (p_2 -
p_1)/(2W)$ giving $\delta \approx 0.08 \pi$. A thick ribbon presentation of the
DNA twist is shown in Fig. \ref{fig:dna}a. 

Our analysis shows that the DNA
surface is fully accessible to a spherical molecule of diameter $D$ if
\begin{equation}
h + D \leq \frac{2}{k} \, \frac{(1 \mp \sin \delta)}{\cos \delta} \ ,
\end{equation}
for the minor ($-$) and major ($+$) grooves, respectively. This gives $D \leq
0.239$ nm for the minor groove and $D \leq 0.795$ nm for the major groove.  In
accord with these estimates, DNA grooves are good binding sites for many ions
and ligands. 
The groove surfaces are fully exposed to water, whose molecular diameter
is about 0.14 nm.  Interestingly, we find that though the groove surface
area varies with $k$ differently for different grooves, the mean surface area
of the two grooves decreases with $k$ (Fig. \ref{fig:dna}b).
By using the total energy given by Eq. (\ref{eq:F}) in the limit of $D=0$, 
for which the exposed area is equal to the total groove surface area,
it is found that the DNA twist is stable at the experimentally observed pitch
for $\kappa \approx 0.0346\,\sigma W^2$ (Fig. \ref{fig:dna}c). 
Fig. \ref{fig:dna}c also shows that the transition from the flat phase
($k=0$) to the twisted phase is discontinuous as the two phases are separated by
an energy barrier. This symmetry breaking is
due to the ribbon's non-zero thickness. We have not considered the full phase
diagram for the DNA ribbon with other types of conformation. However, it can be
expected that for DNA, the elongated twisted conformation is more favorable
than other compact conformations, such as the rolled and the globular ones, due
to strong electrostatic repulsion between DNA charges, which are presumably
located on the ribbon edges. 

In summary, we have shown that a simple surface energy underscoring the
interaction of a ribbon with a solvent may induce dramatic changes
in the ground state conformation of the ribbon, from a planar rolled
conformation to the elongated twisted and the globular shapes.
Such conformational changes and the associated symmetry breaking are moderated
by the ribbon thickness and the solvent molecule's size. Remarkably, the phase
diagram found reflects the dual characters of a ribbon, i.e. membrane-like
as with the rolled conformation, and polymer-like as with the twisted and
the globule conformation. Interesting, the ribbon thickness and the solvent
size are shown to determine this phase behavior.

This research is funded by Vietnam National Foundation for Science and
Technology Development (NAFOSTED) under Grant No. 103.01-2016.61.

\bibliography{refs_ribbon}

\clearpage

\newpage

\setcounter{equation}{0}
\renewcommand\theequation{S\arabic{equation}}

\setcounter{figure}{0}
\renewcommand\thefigure{S\arabic{figure}}

\setcounter{table}{0}
\setcounter{page}{1}

\onecolumngrid

\makeatletter

\section{Supplemental Material: Symmetry breaking in the folding of thick
ribbons moderated by solvent}

\subsection{Differential geometry of thick ribbons}

Consider a ribbon of length $L$, width $W$ and thickness $h$. Assume that the
midplane of the ribbon is differentiable and in Cartesian coordinates it can be
parameterized as
$\vec R_\mathrm{mid} (\alpha,\beta)$, where $\alpha$ and $\beta$ are
the parameters with $\alpha \in [\alpha_0,\alpha_m]$
and $\beta \in [\beta_0, \beta_m]$. The choice of $\alpha$ and $\beta$
will conveniently depend on the type of ribbon conformation and will be given 
in the next subsections. The metric tensor \cite{Kreyszig} of the midplane can
be calculated as \begin{equation}
g_{\alpha \beta} = (\partial_\alpha \vec R_\mathrm{mid}) \cdot
(\partial_\beta \vec R_\mathrm{mid}) \ ,
\end{equation}
where $\partial_\alpha$ and $\partial_\beta$ are partial derivatives with
respect to $\alpha$ and $\beta$, respectively.
The determinant of this tensor is denoted as
\begin{equation}
g = \det(g_{\alpha\beta})\ .
\end{equation}
Assume that the surface area of the midplane is conserved and given by
\begin{equation}
\Sigma = \int_{\alpha_0}^{\alpha_m} d\alpha \int_{\beta_0}^{\beta_m} d\beta\, \sqrt{g} 
= LW \ .
\end{equation}
The local normal vector to the ribbon midplane is given by
\begin{equation}
\hat N =
\frac{(\partial_\alpha \vec R_\mathrm{mid}) \times (\partial_\beta \vec R_\mathrm{mid} )}
{|(\partial_\alpha \vec R_\mathrm{mid}) \times (\partial_\beta \vec R_\mathrm{mid} )|}\ .
\end{equation}
The ribbon physical surfaces can be constructed from the midplane as
\begin{equation}
\vec R_\mathrm{surface}^\pm = \vec R_\mathrm{mid} \pm \frac{h}{2} \hat N \ ,
\label{eq:Rsurf}
\end{equation}
where $\pm$ denote the upper and lower surfaces, respectively. 
One should be able to calculate the metric tensors of these surfaces 
\begin{equation}
g_{\alpha \beta}^\pm = (\partial_\alpha \vec R_\mathrm{surface}^\pm) \cdot
(\partial_\beta \vec R_\mathrm{surface}^\pm) \ ,
\end{equation}
and their determinants
\begin{equation}
g^\pm = \det(g_{\alpha\beta}^\pm)\ .
\end{equation}
The areas of the ribbon upper and lower surfaces are then given by
\begin{equation}
\Sigma^\pm = \int_{\alpha_0}^{\alpha_m} d\alpha \int_{\beta_0}^{\beta_m} d\beta
\, \sqrt{g^\pm} \ .
\end{equation}
Because of the non-zero thickness, $\Sigma^\pm$ are generally different from
$\Sigma$. We will show later that the total surface area of a ribbon needs not
to be conserved, i.e., $\Sigma^{+} + \Sigma^{-} \neq 2 \Sigma$, as found in
the case of the twisted ribbon.

The surface curvatures of the midplane can be obtained by considering the 
second fundamental form tensor \cite{Kreyszig}
\begin{equation}
b_{\alpha \beta} = (\partial_{\alpha,\beta} \vec R_\mathrm{mid} ) \cdot \vec N \ .
\end{equation}
The Gaussian curvature can be calculated as
\begin{equation}
K = \frac{\det(b_{\alpha\beta})}{\det(g_{\alpha\beta})} \ , 
\end{equation}
whereas the mean curvature is given by
\begin{equation}
H = \frac{1}{2} b_{\alpha \beta} g^{\beta \alpha} \ , 
\end{equation}
where $g^{\alpha \beta}$ is the inverse metric tensor
($g_{\alpha \gamma} g^{\gamma \beta} = \delta_{\alpha}^{\beta}$).

\subsection{Exposed surface area and bending energy of the rolled (Archimedean
spiral) ribbon}

We consider a general case of the rolled conformation which has a hole (Fig.
\ref{fig:cartoon}b) in the middle and whose midplane surface is
parameterized as
\begin{equation}
\vec R_\mathrm{mid} (\phi,z) = \left(\frac{\phi}{2\pi}p 
\cos \phi, \frac{\phi}{2\pi}p \sin \phi, z\right) ,
\end{equation}
where 
$\phi \in [\phi_0,\phi_m]$ is the azimuthal angle, 
$p$ is the distance between consecutive turn of the spiral,
and $z \in [-W/2,W/2]$ is the ribbon's lateral coordinate. 
The metric tensor determinant of the midplane is obtained as
\begin{equation}
g = \frac{p^2(1+\phi^2)}{4\pi^2} \ .
\end{equation}
The conservation condition of the midplane surface area is given by
\begin{equation}
\Sigma = LW = \int_{-W/2}^{W/2} dz \, \int_{\phi_0}^{\phi_m} \sqrt{g} \,d\phi
= \left. \frac{W p}{4\pi} \left[\phi \sqrt{1+\phi^2} + \arcsinh(\phi) \right]
\right|_{\phi_0}^{\phi_m} \ ,
\end{equation}
from which one can numerically calculate $\phi_m$ knowing $\phi_0$, $L$, and $p$.
The metric tensor determinants of the ribbon's physical surfaces are given by
\begin{equation}
g^\pm = \left[ \frac{p\sqrt{1+\phi^2}}{2\pi} \pm \frac{h}{2}
\left(\frac{2+\phi^2}{1+\phi^2}\right) \right]^2 \ .
\end{equation}
The surface areas of the ribbon's upper and lower surfaces can be exactly
calculated as
\begin{equation}
\Sigma^{\pm} = \int_{-W/2}^{W/2} dz \int_{\phi_0}^{\phi_m} \sqrt{g^\pm}\, d\phi
= \left. W \left[ \frac{p}{4\pi}\left(\phi \sqrt{1+\phi^2} + \arcsinh(\phi)\right)
\pm \frac{h}{2} \left(\phi + \arctan(\phi)\right) \right] \right|_{\phi_0}^{\phi_m} \ .
\label{eq:sigmarolled}
\end{equation}
As easily seen, the total surface area of the rolled ribbon is conserved, i.e.,
$\Sigma^{+} + \Sigma^{-} = 2 \Sigma$.

For the self-avoidance condition, we will simply assume that $p \ge h$.
For a given solvent diameter $D$, we will consider only the rolled
conformations, such that the contact surfaces between successive turns of the
roll are fully buried, which are found when $p < h + D$.
In these conformations, only the outer surface and
possibly also the inner surface of the spiral are 
exposed. The ribbon's total exposed area thus is given by
\begin{equation}
S_\mathrm{rolled} = W \int_{\phi_m-2\pi}^{\phi_m} \sqrt{g^{+}}\, d\phi
+ W \int_{\phi_0}^{\phi_0+2\pi} \sqrt{g^{-}}\, d\phi \ ,
\label{eq:srolled}
\end{equation}
where the second term in the right hand side is included only if
the hole size is larger than the solvent diameter.
The integrals in Eq. (\ref{eq:srolled}) are easily calculated using
the result of Eq. (\ref{eq:sigmarolled}).

The Gaussian curvature of the midplane surface of the rolled conformation is
always zero, while the mean curvature is obtained as
\begin{equation}
H = - \frac{\pi (2+\phi^2)}{p\, (1+\phi^2)^{3/2}} \ ,
\end{equation}
where the minus sign means that one of the principal curvatures is in 
opposite direction to the normal vector.
For a given $p$, the minimum value of $\phi_0$ can be obtained by requiring
that the spiral radius of curvature at $\phi=\phi_0$ must be larger than $h/2$.
The bending energy can be calculated as
\begin{eqnarray}
U_\mathrm{rolled} & = & \int_{-W/2}^{W/2} dz \int_{\phi_0}^{\phi_m} 
\frac{\kappa}{2} H^2 \sqrt{g}\, d\phi \nonumber \\
& = &
W \frac{\pi \kappa}{4p} \left. \left[
\frac{\phi(9 + 8\phi^2)}{3(1+\phi^2)^{3/2}} + \arcsinh(\phi)
\right] \right|_{\phi_0}^{\phi_m} \ .
\end{eqnarray}

The optimal rolled conformation is obtained by minimizing the total 
energy (Eq. (1)) with respect to $p$ and $\phi_0$. For sufficiently small
stiffness, this minimization leads to a rolled conformation with no hole (Fig.
\ref{fig:cartoon}a) with $p=h$ and $\phi_0\approx 0.541 \pi$. 
For $L \gg h$, one gets
\begin{equation}
\phi_m \approx \left(\frac{4\pi L}{h}\right)^{1/2} \ .
\end{equation}
In this large $L$ limit, the exposed area of the
rolled conformation with no hole can be estimated as 
\begin{equation}
S_\mathrm{rolled} \approx 2 W (\pi h L)^{1/2} \ ,
\end{equation}
whereas the bending energy is approximately given by
\begin{equation}
U_\mathrm{rolled} \approx \frac{\kappa \pi W}{8h} 
 \ln\left( \frac{16 \pi L}{h}\right) \ ,
\end{equation}
given that $\arcsinh(x) = \ln(x + \sqrt{1+x^2})$.

\subsection{Exposed surface area and bending energy of the curled ribbon}

Suppose that the curled ribbon (Fig. \ref{fig:cartoon}c) has $n$ turns.
Denote $r$ the midplane's radius of curvature at the turns. 
We consider only the turns with $h/2 \leq r < (h+D)/2$, so that the ribbon's
inner surface associated with the turns are completely buried.
By simple geometrical consideration,
it can be shown that its exposed surface area for a solvent
diameter $D$ is equal to
\begin{equation}
S_\mathrm{curled} = W \left[ 2 \left(\frac{L - n\pi r}{n+1}\right) 
+ \left(r+\frac{h}{2}\right)\left[n\pi - 2\alpha \Theta(n-1)
- 2 \beta \Theta(n-2)\right] \right] \ ,
\end{equation}
where 
$\alpha=\arcsin(\frac{D}{2r+h+D})$,
$\beta=\arccos(\frac{4r}{2r+h+D})$, 
$\Theta(x)$ is the step function equal to 1 if $x > 0$ and 0 otherwise,
$n$ is an integer satisfying 
$1 \leq n \leq n_\mathrm{max} \approx \frac{L}{\pi r}$. 
The Gaussian curvature of the midplane of the curled ribbon is always zero,
while the mean curvature is non-zero only at the turns, at which $H = 1/r$.
Therefore, the bending energy of the curled ribbon is given by
\begin{equation}
U_\mathrm{curled} = \frac{\kappa n \pi W}{2r} \ .
\end{equation}

The optimal curled conformation is obtained numerically by minimizing the total
energy on changing $r$ and $n$. Suppose that the energy minimum is observed at
$r=r^*$ and $n=n^*$.
The optimal conformation with $n^* = n_\mathrm{max}$ may be obtained for
sufficiently large solvent size ($D \gg h$) and small stiffness. In this
particular conformation, denoted as `crinkled' conformation
\cite{Hoang2012EPL}, the ribbon is globally straight but locally modulated. 
In the limits of small solvent size ($D \ll h$) and 
vanishing stiffness, $n^* \approx \sqrt{n_\mathrm{max}+1}-1$. 
For large length ($L \gg h$), one gets $n^* \approx (2 L/\pi h)^{1/2}$ and 
\begin{equation}
S_\mathrm{curled} \approx 2W (2 \pi h L)^{1/2} \ ,
\end{equation}
\begin{equation}
U_\mathrm{curled} \approx \kappa W h^{-3/2} (2\pi L)^{1/2}\ .
\end{equation}

\subsection{Exposed surface area and bending energy of the twisted ribbon}

We parameterized the midplane of the twisted ribbon (Fig. \ref{fig:cartoon}e) as:
\begin{equation}
\vec R_\mathrm{mid} (u,z) = (u\cos kz, u \sin kz, z) ,
\label{eq:midplane}
\end{equation} 
where $u \in \left[-\frac{W}{2},\frac{W}{2}\right]$, $z \in [0, z_m]$, and $k$
is the wave number of the twisting along the $z$ axis. 
The determinant of the metric tensor of the midplane is given by
\begin{equation}
g = 1 + k^2 u^2 \ .
\end{equation}
$z_\mathrm{m}$ can be determined from the conservation of the midplane surface area:
\begin{equation}
\Sigma = LW = \int_0^{z_\mathrm{m}} dz \int_{-W/2}^{W/2} \sqrt{g} \,du 
=
z_\mathrm{m} \left[\frac{W\sqrt{4+k^2 W^2}}{4} +  
\frac{\arcsinh(\frac{kW}{2})}{k} \right] \ .
\label{eq:zmtwist}
\end{equation}

\begin{figure}
\includegraphics[width=3in]{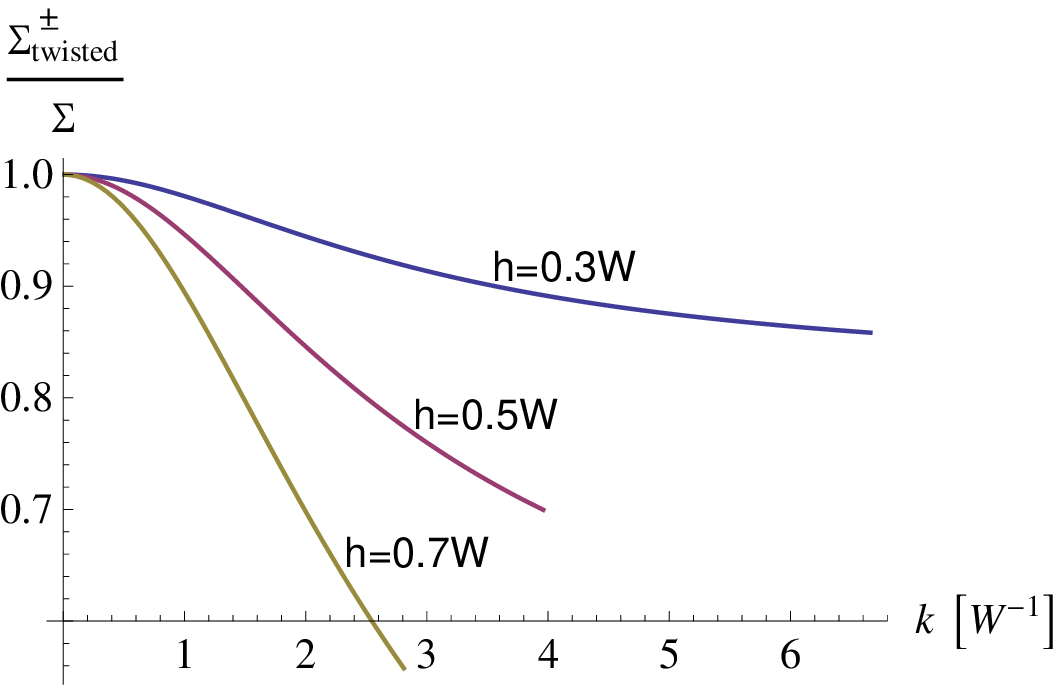}
\caption{
Dependence of the twisted ribbon's upper and lower surface areas,
$\Sigma^{\pm}_\mathrm{twisted}$, on the twist's wave number $k$. The areas are 
shown relative to the midplane area $\Sigma$. The data are obtained
for $h=0.3W$, 0.5$W$ and 0.7$W$ as indicated. The range of $k$ is 
$[0,k_\mathrm{max}]$ with $k_\mathrm{max}=2/h$.
}
\label{fig:stwist}
\end{figure}

The physical surfaces of a twisted ribbon of thickness $h$ can be constructed
from the midplane using the standard procedure as given by Eq. (\ref{eq:Rsurf}). 
The determinants of the metric tensors of the ribbon upper
and lower surfaces are given by
\begin{equation}
g^\pm = \frac{[k^2 h^2 - 4(1 + k^2 u^2)^2]^2}{16(1+k^2u^2)^{3}} \ .
\label{eq:gpm}
\end{equation}
Note that the metric tensor determinant is the same for the
upper and lower surfaces.
The upper and lower surface areas of the twisted ribbon can be
determined analytically and are given by
\begin{equation}
\Sigma_\mathrm{twisted}^\pm =
\int_0^{z_m} dz \int_{-W/2}^{W/2} \sqrt{g^\pm} \, du =
z_m \left[\frac{W\sqrt{4+k^2W^2}}{4} 
+ \frac{\arcsinh(\frac{kW}{2})}{k}
- \frac{h^2 k^2 W}{2\sqrt{4+ k^2 W^2}} \right] \ .
\end{equation}
By using Eq. (\ref{eq:zmtwist}) one obtains
\begin{equation}
\Sigma_\mathrm{twisted}^\pm = \Sigma
- \frac{z_m h^2 k^2 W}{2\sqrt{4+ k^2 W^2}} \ .
\label{eq:sigmatwisted}
\end{equation}
Eq. (\ref{eq:sigmatwisted}) clearly shows that 
the surface areas of the physical surfaces of the twisted ribbon
are smaller than that of the midplane and this is due
to the effect of non-zero thickness $h$.
Fig. \ref{fig:stwist} shows that the ribbon surface area
decreases when either $k$ or $h$ increases. 
Note that $\Sigma_\mathrm{twisted}^\pm$ is also the exposed area
for the case of $D=0$, thus the twisted conformation is
favorable in terms of surface energy for any solvent size. We will
show that this is also true for the case of DNA twist later
in this supplemental material.

The ribbon thickness also has a strong effect on self-avoidance condition of
the twisted conformation. If the ribbon has zero thickness, the twist's wave
number $k$ can increase to infinity while the ribbon size along
the $z$ axis shrinks to zero. If the ribbon has a finite thickness,
self-avoidance prevents infinite twisting. As the wave number $k$ increases to
a certain value, the physical surface of the ribbon starts to intersect itself.
The onset of self-intersection occurs exactly at the point where the metric of the
surface vanishes ($g^\pm=0$) and ought to happen at $u=0$ due to
symmetry. By using Eq. (\ref{eq:gpm}) for $g^\pm$, one obtains the
self-avoidance condition as
\begin{equation}
k \leq k_\mathrm{SA} = \frac{2}{h} \ ,
\end{equation}
where $k_\mathrm{SA}$ is the self-avoidance limit of $k$ determined by the thickness
$h$.

For a twisted ribbon submerged in a solvent, the closest distance from the
center of a solvent molecule to the ribbon midplane is $(h+D)/2$. One can
construct the excluded volume surfaces as
\begin{equation}
\vec R_\mathrm{ES}^\pm = \vec R_\mathrm{mid} \pm \frac{h+D}{2} \hat N .
\end{equation}
These new surfaces, unlike the ribbon’s physical surfaces, are virtual surfaces
and can self-intersect. This self-intersection indicates that some 
regions of the ribbon physical surfaces are inaccessible to the solvent. 
Similarly to the self-avoidance condition, the self-intersection happens only
when $k$ is sufficiently large, i.e.
\begin{equation}
k \geq k_\mathrm{D} = \frac{2}{h+D} \ .
\end{equation}
The intersection line of the excluded volume surface corresponds to the borders
of the buried area on the physical surface. For the twisted ribbon, it 
is a helical curve lying midway between the ribbon's successive turns and has a
constant $u$ coordinate.

\begin{figure}
\includegraphics[width=3in]{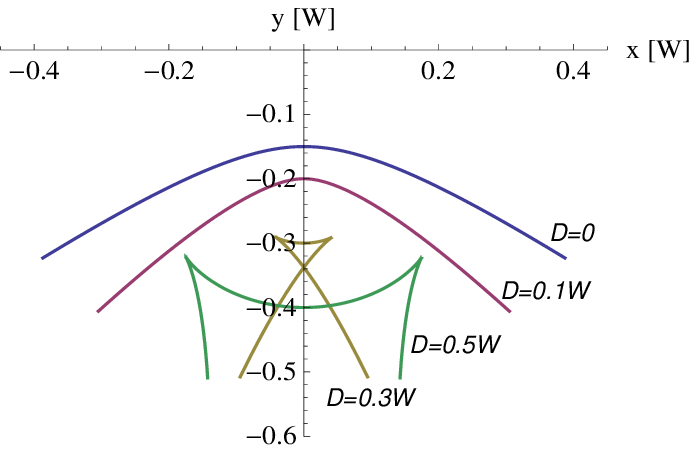}
\caption{
Contours of the intersections of the excluded volume surface of a
twisted ribbon with the $z=0$ plane for four solvent diameters,
$D=0$, $0.1W$, $0.3W$ and $0.5 W$, as indicated.
The contours are calculated for a twisted ribbon of width $W=1$, thickness
$h=0.3 W$ and wave number $k=(4\pi/3) W^{-1}$. 
The self-intersection of the contour on the $y$ axis is seen for $D=0.3 W$.
}
\label{fig:contour}
\end{figure}

In order to calculate the ribbon's exposed area, we first determine the $u$
coordinate of the excluded volume surface intersection.
Thanks to symmetry, this task can be done by considering the intersection
contour of one of the excluded volume surface with the $z=0$ plane. For the
upper excluded volume surface, the contour's coordinates can be found as
\begin{equation}
\vec C^{+} (u) \equiv 
\left. \vec R_{ES}^{+} \right|_{z=0} = 
\left[
u\cos(k^2 u v) - v \sin(k^2 u v), -v \cos(k^2 u v) - u \sin(k^2 u v), 0
\right]
\end{equation}
\begin{equation}
\mathrm{with} \qquad v=\frac{(h+D)}{\sqrt{1+k^2 u^2}} \ .
\end{equation}
The contour $\vec C^{-}(u)$ corresponding to the lower excluded volume surface can be obtained
from the above equations by just changing the sign of $v$.
Fig. \ref{fig:contour} shows that the self-intersection of the contour
occurs only on the $y$ axis of the $z=0$ plane. Thus,
the $u$ coordinate of the self-intersection
is the solution of the following equation
\begin{equation}
 \tan \left(\frac{k^2u(h+D)}{2\sqrt{k^2 u^2+1}}\right) = 
\frac{2 u \sqrt{k^2u^2+1}}{h+D} \ ,
\label{eq:tan}
\end{equation}
which can be solved numerically. 
We are interested in only the solution $0< u^* \leq \frac{W}{2}$,
given that $-u^*$ is another solution by symmetry.
For a given $k$, such that  $k_D \le k \le k_{SA}$, and a solvent size $D$,
solving this equation leads to one of the two following situations: (a) there is
a solution $0 < u^* \leq \frac{W}{2}$ (the ribbon surfaces are partially
exposed), and (b) there is no such solution (the ribbon surfaces are
completely shielded). 
In case a, the ribbon's total exposed area on both upper and lower
surfaces is given by:
\begin{eqnarray}
S_\mathrm{twisted} & = & 
4 \int_{0}^{z_m} dz \int_{u^*}^{\frac{W}{2}} \sqrt {g^\pm} \; du \nonumber \\
& = &
\left. 2\, z_m \left( u \sqrt{1+ k^2 u^2} + \frac{\arcsinh(ku)}{k}
-\frac{h^2 k^2 u}{2\sqrt{1 + k^2 u^2}} \right) \right|_{u^*}^{W/2} \ .
\end{eqnarray}
The exposed area is equal to zero for $D > D^*$, where $D^*$ is
a solvent diameter such that $u^*=W/2$. The value of $D^*$ can be determined
numerically through Eq. (\ref{eq:tan}). It can also be shown that
\begin{equation}
h + D^* < \frac{\pi}{k}\sqrt{1+(2/kW)^2} \ .
\end{equation}

The midplane of the twisted conformation has a zero mean curvature, $H=0$, as
found for ideal helicoid \cite{Kreyszig}. The Gaussian curvature can be
determined through calculating the second fundamental form tensor and is given
by
\begin{equation}
K = - \frac{k^2}{(1+k^2 u^2)^2} \ .
\end{equation}
The bending energy for the twisted configuration is calculated as:
\begin{eqnarray}
U_\mathrm{twisted} &=& \iint (-\kappa K) \sqrt{g} \; du\, dz \\
&=&
\int_0^{z_\mathrm{m}} dz \int_{-W/2}^{W/2} \frac{\kappa k^2}{(1+k^2 u^2)^{3/2}} \, du \\
&=&
\kappa z_\mathrm{m} \frac{2 k^2 W}{\sqrt{4 + k^2 W^2}} \ .
\end{eqnarray}

Because $z_m$ is approximately a linear function of $L$, for $L \gg W$ one
find that, for the twisted ribbon, $S_\mathrm{twisted} \propto L$ and
$U_\mathrm{twisted} \propto L$.

\subsection{The spherical spiral (globular) ribbon}

\begin{figure}
\includegraphics[width=0.5\columnwidth]{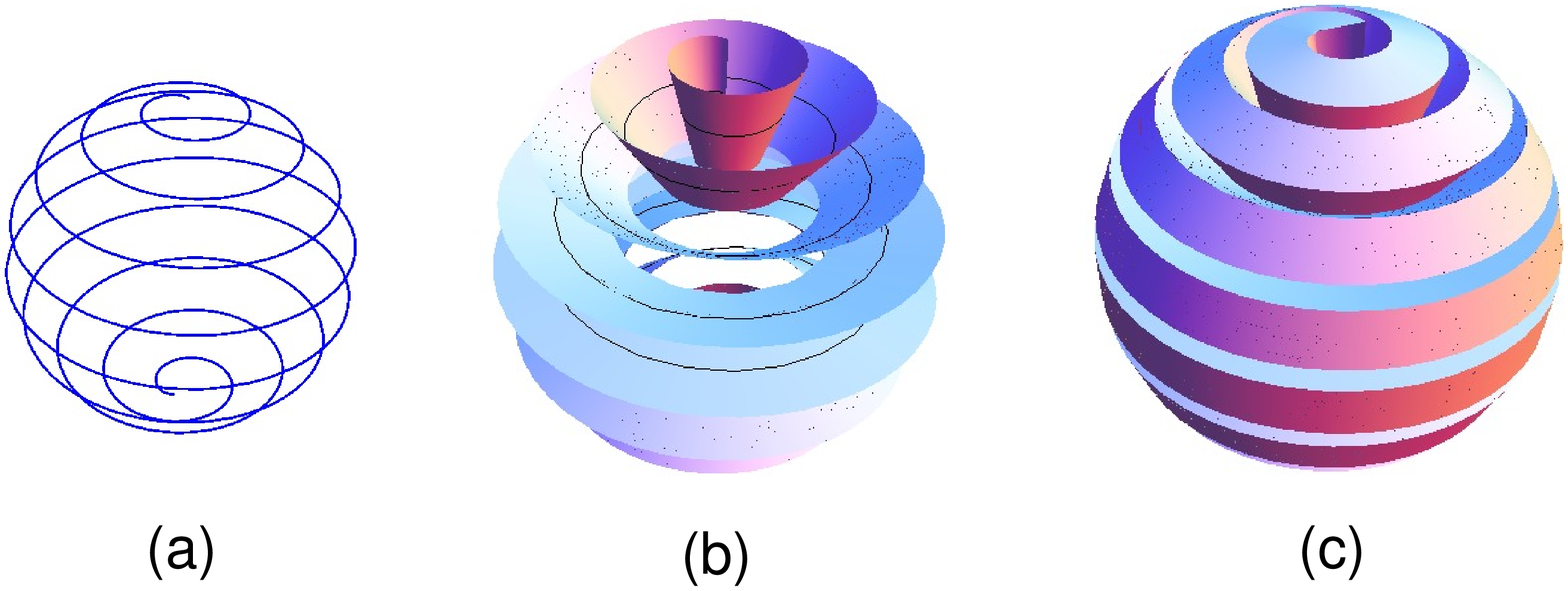}
\caption{
(a) A spherical spiral curve.
(b) The midplane of a spherical spiral ribbon 
with the ribbon center line following the curve shown in a.
(c) A thick spherical spiral (globular) ribbon with the midplane shown in b.
}
\label{fig:globule}
\end{figure}

Our simulations show that for sufficiently large $L$ and sufficiently large
$D$, a ribbon may form compact conformations close to a globular shape with
a small exposed area of its hydrophobic surface. In these globule-like
conformations, the non-hydrophobic edges of the ribbon are exposed
to the solvent. These conformations also display a significant helical feature.
Some non-regularities observed may be due to the small sizes of the
ribbons. There are many possible conformations of a globular ribbon. Here,
based on the hints from simulation result, we consider a parameterized model
with a globular shape, in which the ribbon
forms a {\it spherical spiral} as shown in Fig. \ref{fig:globule}.
The spherical spiral has two poles corresponding to the two ends of the ribbon.
Overall, it forms a spherical layer of thickness $W$. 
The midplane of the spherical spiral ribbon can be parameterized as
\begin{equation}
\vec R_\mathrm{mid} (\phi,u) = \left[ (R+u) \sin(k \phi)
\cos \phi, (R+u)\sin(k \phi) \sin \phi,
(R+u) \cos(k \phi) \right] \ ,
\end{equation}
where $R$ is the radius of the sphere passing through the central curve of the
ribbon, $u \in [-W/2,W/2]$, $\phi \in [0,\phi_m]$ and $k=\pi/\phi_m$.
It is straightforward to calculate the metric tensor of the midplane, whose
determinant is given by 
\begin{equation}
g = (R+u)^2[k^2+\sin^2(k\phi)] \ .
\end{equation}
For a given $\phi_m$, and $k=\pi/\phi_m$, the radius $R$ can be determined from
the midplane surface area conservation
\begin{equation}
\Sigma = LW = \int_{-W/2}^{W/2} du \int_0^{\phi_m} \sqrt{g} \, d\phi 
= R W \int_0^{\phi_m} 
\sqrt{k^2+ \sin^2(k\phi)} \,d\phi \ .
\end{equation}
The upper and lower surfaces of the ribbon can be constructed from the
midplane as shown in Fig. \ref{fig:globule}. The metric tensor determinants
of these surfaces are given by
\begin{equation}
g^\pm = \left[(R+u)\sqrt{k^2+\sin^2(k\phi)} \pm h \cos(k\phi)
\frac{2 k^2 + \sin^2(k\phi)}{2 k^2 + 2 \sin^2(k\phi)}
\right]^2 \ .
\end{equation}
It can be easily shown that $\Sigma^{+} = \Sigma^{-} = \Sigma$.

To check the self-avoidance condition and to calculate the exposed area, we
will employ an approximate approach by considering the cross section of the
ribbon with the sphere of radius $(R+u)$. Such a cross section is a spherical
spiral stripe of the width equal to 
\begin{equation}
{\cal W} (u,h) = 2(R+u)\arctan\left(\frac{h}{2(R+u)}\right) \ ,
\end{equation}
and the midline contour length equal to
\begin{equation}
{\cal L}(u) = \int_0^{\phi_m} |\partial_\phi \vec R_\mathrm{mid}|\, d\phi 
= (R+u) \int_0^{\phi_m} \sqrt{k^2 + \sin^2(k\phi)} \, d\phi
= \frac{L(R+u)}{R} \ .
\end{equation}
For a self-avoiding ribbon, the area of the stripe must be not larger
than the surface area of the sphere
\begin{equation}
{\cal L}(u) \cdot {\cal W}(u,h) \leq 4\pi (R+u)^2 \quad \Rightarrow
\quad
\arctan\left(\frac{h}{2(R+u)}\right) \leq \frac{2\pi R}{L} \ . 
\end{equation}
It is enough to check the above inequality for $u=-W/2$.

By using the same argument as above for a stripe corresponding to 
the excluded volume surface of the ribbon with a solvent of diameter $D$, one
finds that the ribbon surface elements at a given parameter $u$ is exposed to
the solvent if 
\begin{equation}
\arctan\left(\frac{h+D}{2(R+u)}\right) \leq \frac{2\pi R}{L} \ . 
\end{equation}
Assume that the equality of the above equation is found for $u^*$, such that
$-\frac{W}{2} \leq u^* \leq \frac{W}{2}$, the ribbon's exposed area is given by
\begin{equation}
S_\mathrm{globule} = 2 \int_{0}^{\phi_m} d\phi 
\int_{u^*}^{W/2} \sqrt{g^{+}} \, du 
= \frac{L}{R}\left(2R + \frac{W}{2} + u^*\right)\left(\frac{W}{2}-u^*\right)
\ .
\end{equation}

\begin{figure}
\includegraphics[width=3in]{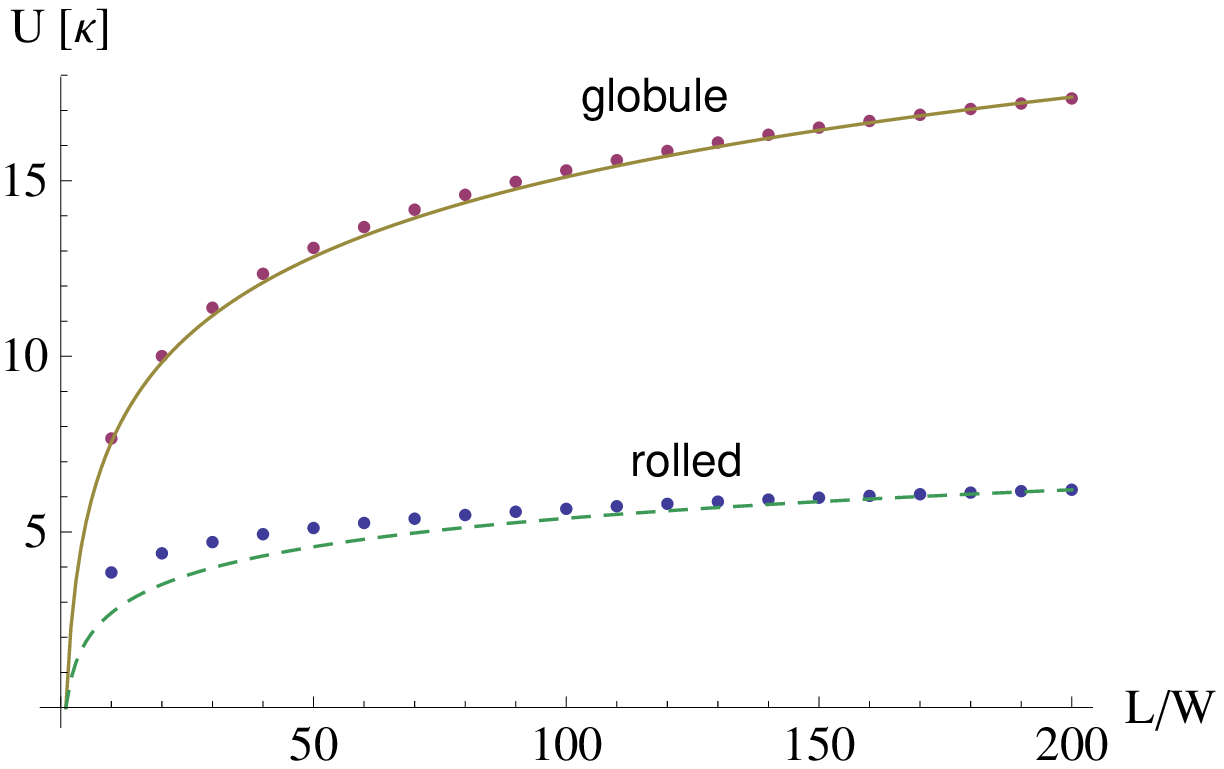}
\caption{
Dependence of bending energy, $U$, on length, $L$, of the rolled ribbon with no
hole and the spherical spiral (globular) ribbon. Discrete points are
obtained by numerical evaluation, whereas smooth curves represent fits with the
$\log(L)$ dependence. The data are obtained for tightly folded ribbons of
thickness $h=0.5W$.
}
\label{fig:uglob}
\end{figure}

By calculating the second fundamental form tensor of the midplane, one
immediately finds that the spherical spiral ribbon has a zero
Gaussian curvature, $K=0$,
whereas the mean curvature is given by
\begin{equation}
H = -\frac{\cos(k\phi)[2k^2+\sin^2(k\phi)]}
{2(R+u)[k^2+\sin^2(k\phi)]^{3/2}} \ .
\end{equation}
The bending energy thus can be numerically calculated from the integral
\begin{eqnarray}
U_\mathrm{globule} &=& \int_0^{\phi_m} d\phi \int_{-W/2}^{W/2} du \;
\frac{\kappa}{2} H^2 \sqrt{g} \nonumber \\
&=& \frac{\kappa}{8} \ln\left(\frac{2R+W}{2R-W}\right)
\int_0^{\phi_m} \frac{\cos^2(k\phi)[2k^2+\sin^2(k\phi)]^2}
{[k^2+\sin^2(k\phi)]^{5/2}} \, d\phi
\ .
\end{eqnarray}
Note that $R$ also depends on $\phi_m$. Our numerical calculations indicate
that the bending energy of the tightly folded spherical spiral ribbon grows
logarithmically with $L$, similar to that of the rolled conformation (Fig.
\ref{fig:uglob}). It is also shown that $U_\mathrm{globule}$ is larger than
$U_\mathrm{rolled}$.

The optimal spherical spiral conformation is obtained by minimizing the
total energy on changing $\phi_m$. In the large length limit ($L \gg h$), one
can write 
\begin{equation}
S_\mathrm{globule} \propto L \qquad \mathrm{and} \qquad
U_\mathrm{globule} \propto \frac{W}{h} \ln L \ .
\end{equation}

\subsection{Ground state phase diagram of thick ribbon}

\begin{figure}
\includegraphics[width=3.3in]{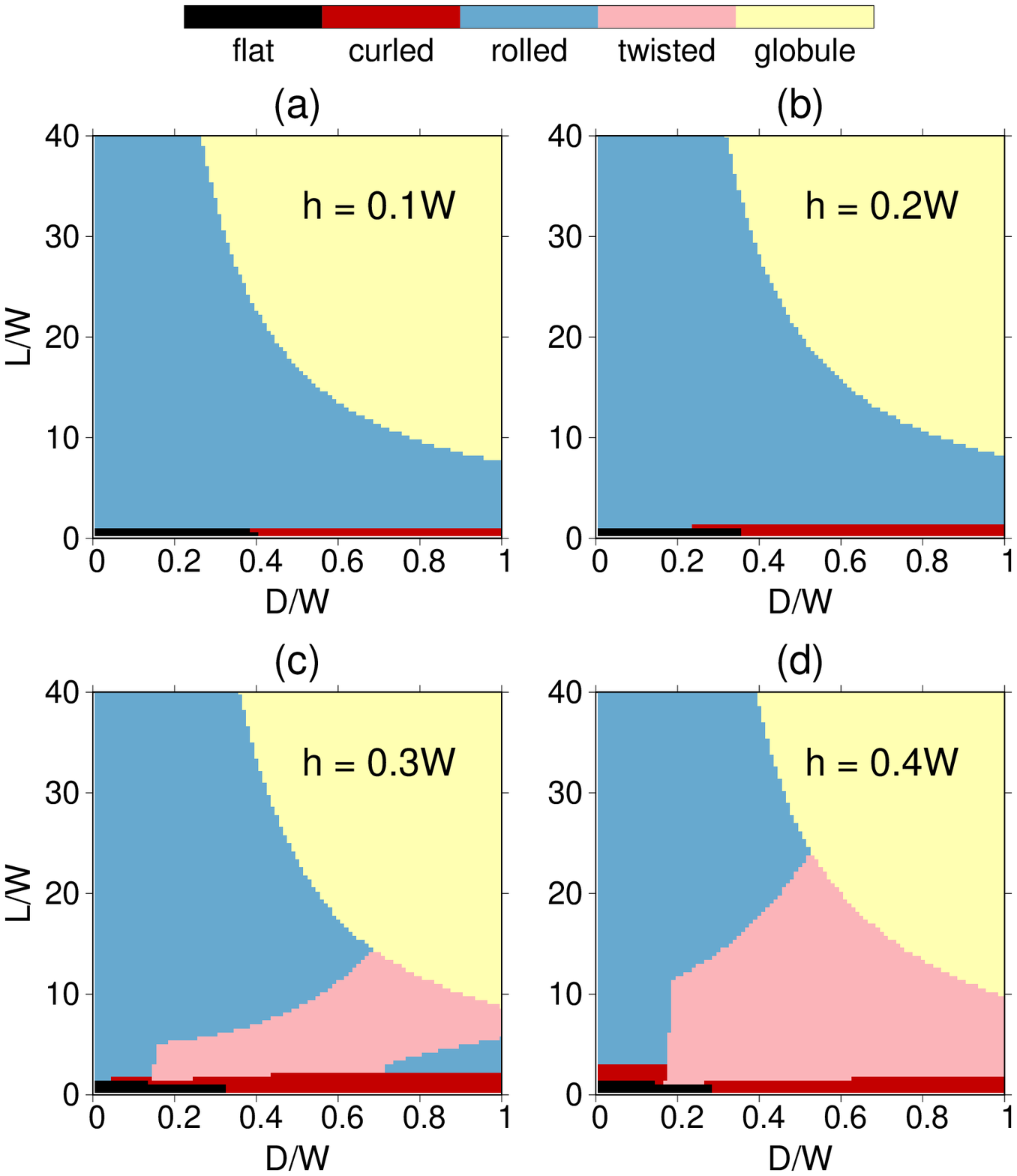}
\caption{
Ground state phase diagram of ribbons as function of the ribbon's length $L$
and solvent diameter $D$. The phase diagram is shown for different ribbon's
thicknesses, $h=0.1W$ (a), $0.2W$ (b), $0.3W$ (c) and $0.4W$ (d).
The bending stiffness is $\kappa=0.1\,\sigma W^2$ for all cases. 
Different phases are indicated by colors as given in the legends (top).
}
\label{fig:phaseh}
\end{figure}

We studied the ground state phase diagram of thick ribbon as function of
the ribbon's length $L$ and the solvent diameter $D$. For a ribbon 
of given $L$ and $D$, together with the thickness $h$ and 
stiffness $\kappa$, each of the four conformations, the rolled, the
curled, the twisted and the spherical spiral ones, is optimized in
terms of their total energies. The ground state is the lowest energy
conformation among the four optimized configurations.
Fig. \ref{fig:diagram} in the main text shows the phase diagrams for
ribbons of the same thickness $h$ but for different values of
stiffness $\kappa$. On the other hand, Fig. \ref{fig:phaseh} shows
the phase diagrams for ribbons of the same $\kappa$ but 
different thicknesses $h$. It is shown that the twisted conformation
appears as the ground state for ribbons of either low stiffness (as shown in
Fig.  \ref{fig:diagram}) or large thickness (Fig. \ref{fig:phaseh}).

\subsection{Thick ribbon description of DNA twist}

Consider the B-DNA double helix structure as a twisted thick ribbon.
We parameterize the midplane of this ribbon as
\begin{equation}
{\vec R}_\mathrm{mid} (u,z) = (u \cos \delta \cos kz, 
u \cos \delta \sin kz, z + u \sin \delta) ,
\end{equation}
where $u \in [-W/2,W/2]$ and $k=2\pi/p$ with $W$ and $p$ correspond to the
width and the pitch of DNA, respectively, $\delta$ is the tilt angle of the
ribbon's lateral direction with respect to the plane perpendicular to the main
axis of the twist (the $z$ axis). The ribbon thickness is denoted $h$.
One can calculate the metric tensor of the midplane and obtain the
metric tensor determinant
\begin{equation}
g 
= (1 + k^2 u^2) \cos^2 \delta \ .
\end{equation}

Denote $h$ the thickness of the DNA ribbon. It is straightforward to construct
the ribbon upper and lower physical surfaces ${\vec R}^{\pm}_\mathrm{surface}$.
These surfaces are shown in Fig. \ref{fig:dna} for realistic parameters of DNA
with a clear appearance of the minor ($+$) and major ($-$) grooves. The metric
tensor determinants of the physical surfaces are given by
\begin{eqnarray}
g^{\pm} & = & 
\frac{\left[
(h^2 k^2 - 4(1+k^2u^2)^2) \cos\delta \pm 
2hk(2+k^2u^2)\sqrt{1+k^2u^2} \sin \delta
\right]^2}
{16 (1+k^2 u^2)^{3}} \ .
\end{eqnarray}
Note that due to the tilt angle $\delta$, the obtained metric tensors are
different for the upper and lower surfaces.
Thus, the surface areas of the grooves are also different as calculated by
\begin{equation}
\Sigma^\pm = \int_0^{z_m} dz \int_{-W/2}^{W/2} du \, \sqrt{g^\pm} \ .
\end{equation}
In fact, for $k>0$, it is found that that $\Sigma^{+} < \Sigma < \Sigma^{-}$ and
$\Sigma^{+} + \Sigma^{-} < 2 \Sigma$ with $\Sigma$ the midplane area. 
Fig. \ref{fig:dna}b shows that the total groove surface area 
decreases with $k$.

If $k$ is increased, the ribbon surfaces can intersect themselves. Like for the
case of the ideal twisted ribbon, the self-intersection starts at
$u=0$ first with a vanishing metric at that point. The latter yields the
self-avoidance condition of the ribbon as
\begin{equation}
h \leq \frac{2}{k} \, \frac{(1 \mp \sin \delta)}{\cos \delta} \ ,
\end{equation}
for the minor ($-$) and major ($+$) grooves, respectively. Because the minor
groove yields a smaller limit for $h$, the self-avoidance constraint is imposed
by the minor groove.

\begin{figure}
\includegraphics[width=3in]{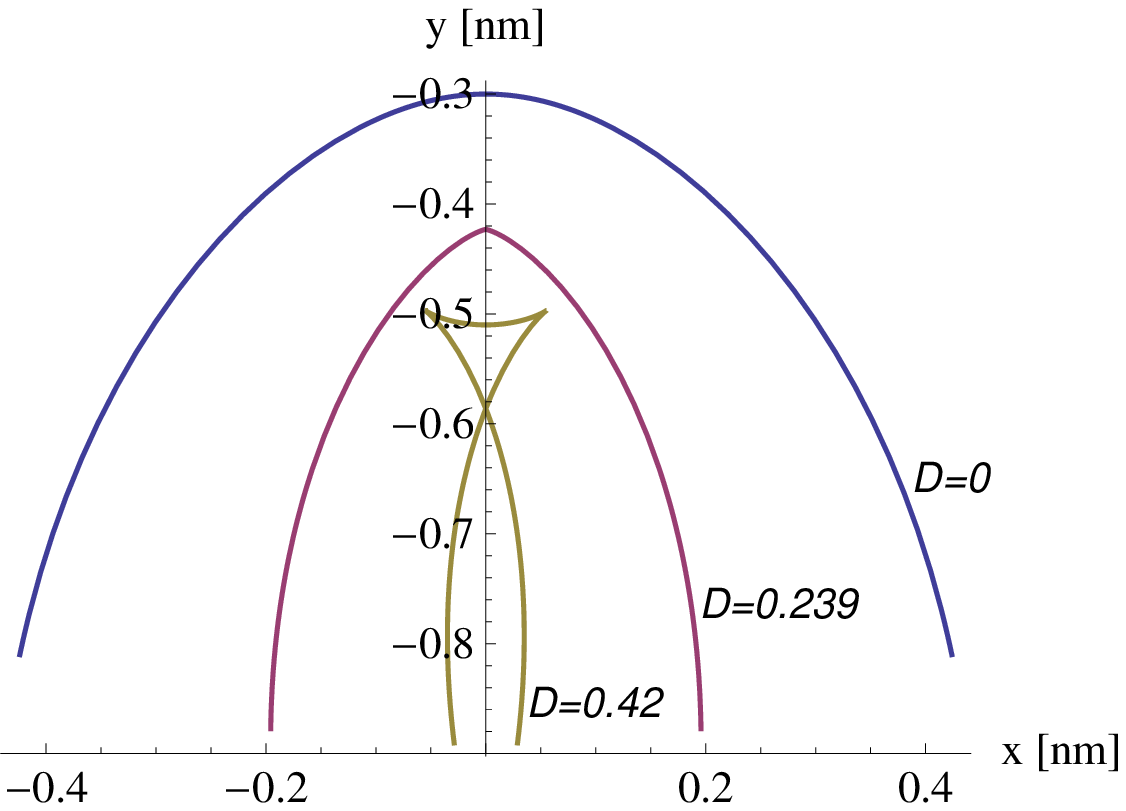}
\caption{
Contours of the intersections of the excluded volume surface, ${\vec
R}_{ES}^{+}$, of the DNA minor groove with the $z=0$ plane for three 
solvent diameters, $D=0$, 0.239 nm and 0.42 nm, as indicated. The contours are
calculated by using realistic DNA parameters of $W=2$ nm, $h=0.6$ nm, $p=3.4$
nm and $\delta=0.08\pi$. 
The case of $D=0.239$ nm corresponds to the contour that is about to
intersect itself at $u=0$.
For $D=0.42$ nm, the self-intersection of the contour is seen on
$y$ axis.
}
\label{fig:dnaminor}
\end{figure}

\begin{figure}
\includegraphics[width=3in]{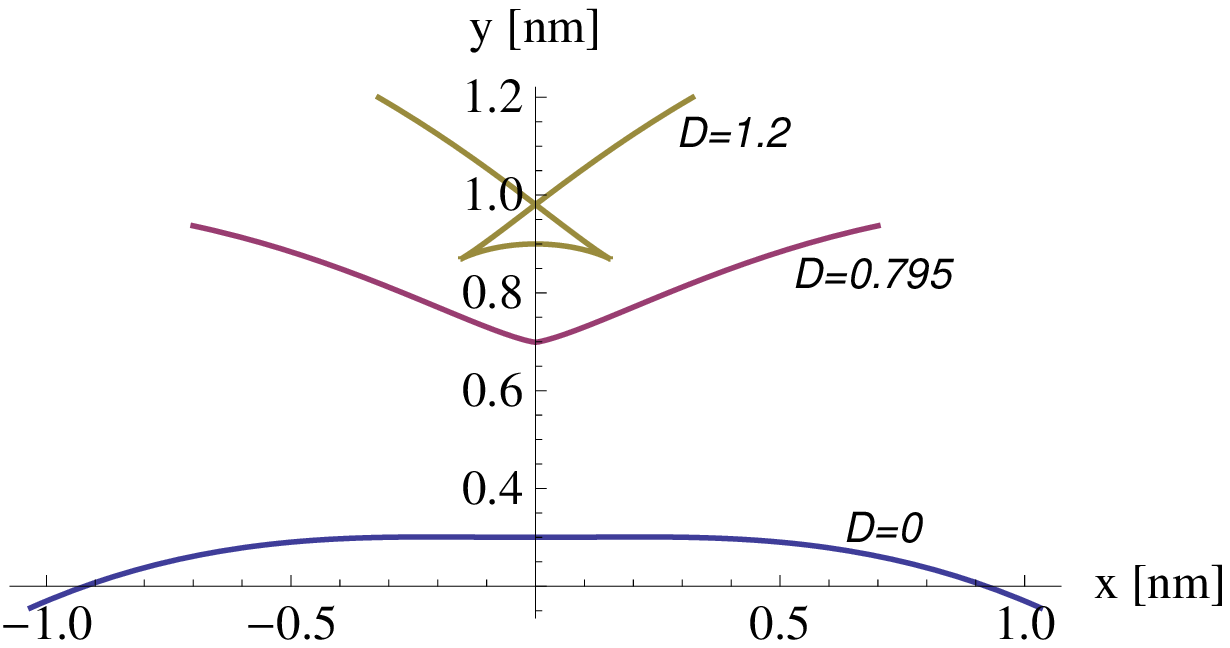}
\caption{
Same as Fig. \ref{fig:dnaminor} but for the DNA major groove and for three
values of solvent diameter, $D=0$, 0.795 nm and 1.2 nm, as indicated.
The case of $D=0.759$ nm corresponds to the contour that is about to
intersect itself at $u=0$.
}
\label{fig:dnamajor}
\end{figure}

\begin{figure}
\includegraphics[width=3in]{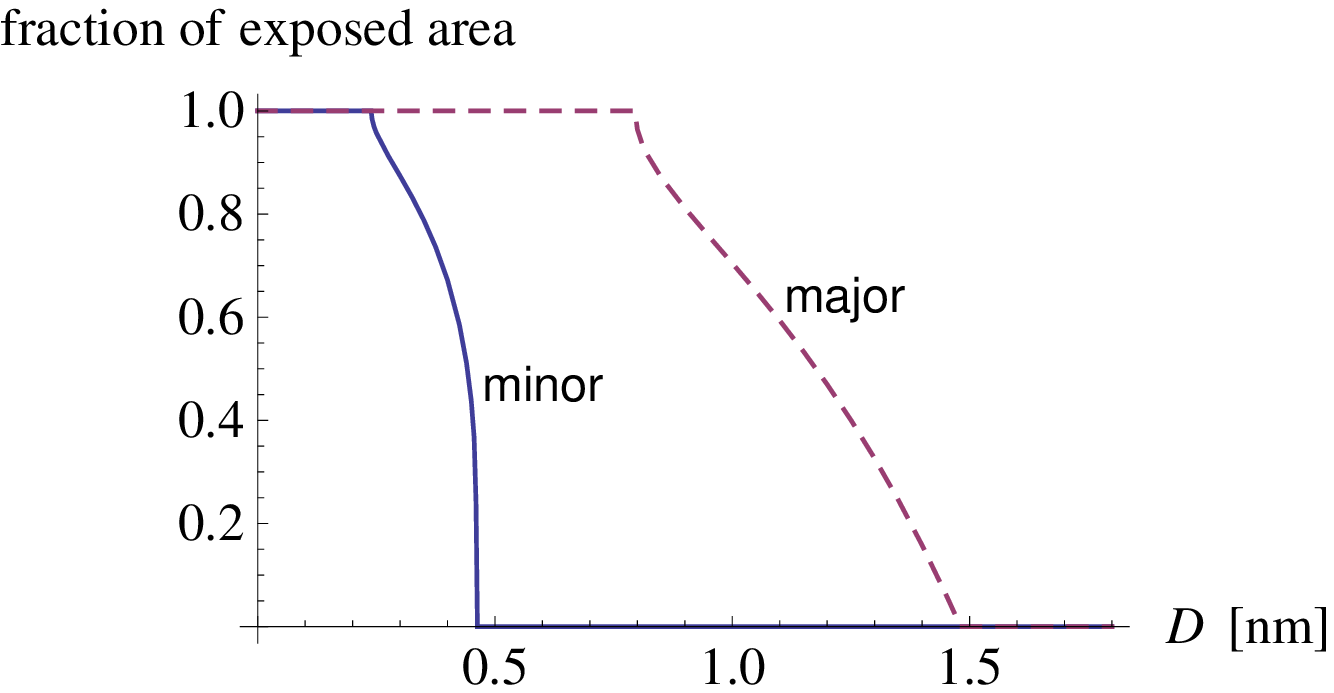}
\caption{
Dependence of the fraction of exposed area on the solvent diameter $D$
for the DNA minor (solid) and major (dashed) grooves. The data are obtained
by using realistic parameters for DNA as given in the caption of Fig.
\ref{fig:dnaminor}.
}
\label{fig:dnaexp}
\end{figure}

In order to calculate the exposed area of the ribbon for a solvent of diameter
$D$, one constructs the ribbon's excluded volume surfaces $\vec R^{\pm}_{ES}$.
The latter have the same form as $\vec R_\mathrm{surface}^\pm$ with $h$ being
replaced by $h+D$. The ribbon surface is fully exposed to solvent if the
excluded volume surface does not self-intersect, which means that
\begin{equation}
h + D \leq \frac{2}{k} \, \frac{(1 \mp \sin \delta)}{\cos \delta} \ ,
\end{equation}
for the minor ($-$) and major ($+$) grooves, respectively.
If the excluded volume surface of a groove self-intersects, the groove surface
is partially exposed. Like for the case of the ideal twisted ribbon, the
$u^*$ position of the self-intersection can be determined numerically by
considering the contour of the excluded volume surface
${\vec R}_{ES}^{\pm}$ on the $z=0$ plane. Suppose that we consider only
$0 < u^* \leq W/2$.
Figs. \ref{fig:dnaminor} \& \ref{fig:dnamajor} show that the contour intersects
itself always on the $y$ axis and this starts happen at a lower solvent size
for the minor groove. Thus, the two grooves can have different values of $u^*$,
denoted as $u^*_{+}$ and $u^*_{-}$.  
The exposed areas of the grooves are given by
\begin{equation}
S_\mathrm{DNA}^{\pm} = 2 \int_0^{z_m} dz \int_{u^*_{\pm}}^{W/2} \sqrt{g^\pm} \, du \ .
\end{equation}
Fig. \ref{fig:dnaexp} shows the dependence of the fraction of exposed area 
on the solvent diameter $D$ for the two grooves.

The curvatures of the midplane of the DNA ribbon can be determined by
calculating the tensors of the first and second fundamental forms.
One obtains the mean curvature
\begin{equation}
H = \frac{k(-2 + k^2 u^2 + k^4 u^4 \cos^2 \delta) \sin(2\delta)}
{4\sqrt{1+ k^2 u^2}} \ ,
\end{equation}
and the Gaussian curvature
\begin{equation}
K = -\frac{k^2}{(1+k^2 u^2)^2} \ .
\end{equation}
Note that unlike the ideal helicoid, the mean curvature of the DNA midplane
is mostly non-zero for $\delta \ne 0$. The Gaussian curvature of the DNA
ribbon, on the other hand, remains the same as for the ideal helicoid. The
bending energy of the DNA ribbon can be calculated numerically by integrating
the bending energy density over the midplane surface 
\begin{equation}
U_\mathrm{DNA} = \frac{\kappa}{2} \int_0^{z_m} dz \int_{-W/2}^{W/2} 
 (H^2 - 2 K) \sqrt{g} \, du \ .
\end{equation}

\end{document}